\documentclass[aps,preprint,amsmath,amssymb,floatfix]{revtex4}
\usepackage{graphicx,morefloats}
\usepackage{color}

\begin{document}
\hyphenation{va-ni-sh-ing}

\begin{center}
{\large\bf Destruction of Kondo effect in cubic heavy fermion compound
Ce$_3$Pd$_{20}$Si$_6$}
\\[0.5cm]

J.\ Custers$^1$, K.-A.\ Lorenzer$^1$, M.\ M\"uller$^1$, A.\ Prokofiev$^1$, A.\
Sidorenko$^1$, H.\ Winkler$^1$, A.\ M.\ Strydom$^2$, Y.\ Shimura$^3$, T.\
Sakakibara$^3$, R.\ Yu$^4$, Q.\ Si$^4$, and S.\ Paschen$^{1\ast}$\\

$^1$ Institute of Solid State Physics, Vienna University of Technology, 1040
Vienna, Austria\\

$^2$ Physics Department, University of Johannesburg, Auckland Park 2006, South
Africa\\

$^3$ Institute for Solid State Physics, University of Tokyo, Kashiwa 277-8581,
Japan\\

$^4$ Department of Physics and Astronomy, Rice University, Houston, TX 77005,
USA\\

$^{\ast}$ e-mail: paschen@ifp.tuwien.ac.at

\end{center}

\vspace{0.5cm}
{\bf How ground states of quantum matter transform between one another reveals
deep insights into the mechanisms stabilizing them. Correspondingly, quantum
phase transitions are explored in numerous materials classes, with heavy fermion
compounds being among the most prominent ones. Recent studies in an anisotropic
heavy fermion compound have shown that different types of transitions are
induced by variations of chemical or  external pressure
\cite{Fri09.1,Tok09.1,Cus10.1}, raising the question of the extent to which
heavy fermion quantum criticality is universal. To make progress, it is
essential to broaden both the materials basis and the microscopic parameter
variety. Here, we identify a cubic heavy fermion material as exhibiting a
field-induced quantum phase transition, and show how the material can be used
to explore one extreme of the dimensionality axis. The transition between two
different ordered phases is accompanied by an abrupt change of Fermi surface,
reminiscent of what happens across the field-induced antiferromagnetic to
paramagnetic transition in the anisotropic YbRh$_2$Si$_2$. This finding leads to
a materials-based global phase diagram -- a precondition for a unified
theoretical description.}

\newpage

Quantum phase transitions arise in matter at zero temperature due to competing
interactions. When they are continuous, the associated quantum critical points
(QCPs) give rise to collective excitations which influence the physical
properties over a wide range of parameters. As such, they are being explored in
a variety of electronic materials, ranging from high $T_{\mathrm{c}}$ cuprates
to insulating magnets and quantum Hall systems \cite{Sch10.1,Bro08.1}.

Heavy fermion compounds are prototype materials to study quantum phase
transitions. Their low energy scales allow to induce such transitions
deliberately, by the variation of external parameters such as pressure or
magnetic field. Microscopically, electrons in partially-filled $f$ shells behave
as localized magnetic moments. They interact with conduction electrons through a
Kondo exchange interaction, which favors a non-magnetic ground state that
entangles the local moments and the spins of the conduction electrons. They also
interact among themselves through an RKKY exchange interaction, which typically
induces antiferromagnetic order. The best characterized  QCPs occur in heavy
fermion compounds with anisotropic structures. Examples include the monoclinic
CeCu$_{\mathrm{6-x}}$Au$_{\mathrm{x}}$ (ref.\,\onlinecite{Loe94.1}), and
tetragonal CePd$_2$Si$_2$ (ref.\,\onlinecite{Mat98.1}), YbRh$_2$Si$_2$
(ref.\,\onlinecite{Geg07.1}) and Ce$M$In$_5$ ($M$=Co,Rh) \cite{Par06.1}.

It has been known that tuning external parameters changes the ratio of the Kondo
coupling to the RKKY interaction. Recently, the importance of a second
microscopic quantity has been suggested. This is the degree of quantum
fluctuations of the local moments, parameterized by $G$: magnetic order weakens
with increasing $G$, as it would with enhancing the Kondo coupling
$J_{\mathrm{K}}$. These two quantities define a two-dimensional parameter space,
which allows the consideration of a global phase diagram \cite{Si06.1}. This
global phase diagram is most clearly specified via the energy scale $T^{\ast}$
associated with the breakdown of the Kondo entanglement between the local moments
and conduction electrons. So far $T^{\ast}$ has only been identified in
tetragonal YbRh$_2$Si$_2$ (refs.\,\onlinecite{Pas04.1, Geg07.1,Fri10.2}). It is
believed that this energy scale not only provides a general characterization of
the heavy fermion quantum criticality but also underlies the non-Fermi liquid
behaviour and anomalous dynamical scaling \cite{Aro95.1,Sch00.1,Fri10.2} observed
in these and related materials. 

In order to explore the additional axis of the global phase diagram, we take
advantage of the fact that enhancing spatial dimensionality reduces the quantum
fluctuation parameter $G$. Therefore, it would be invalubale to study the extreme
three-dimensional cubic heavy fermion compounds and compare their quantum
critical behaviour with that of the tetragonal and other non-cubic  materials.

Here, we do so in the cubic heavy fermion compound Ce$_3$Pd$_{20}$Si$_6$. We
show that this material undergoes a quantum phase transition at a readily
accessible magnetic field, and are able to identify the Kondo breakdown energy
scale $T^{\ast}$. This scale vanishes inside the ordered part of the compound's
phase diagram, thereby providing the first clear evidence for a Kondo breakdown
in the three-dimensional part of the global phase diagram. Furthermore, we
show that this vanishing scale is the origin of the non-Fermi liquid behaviour
observed previously in this compound (Supplementary Information).

In the cubic crystal structure of space group $Fm\bar{3}m$, the Ce atoms occupy
two different crystallographic sites, $4a$ (Ce1) and $8c$ (Ce2), both with cubic
point symmetry~\cite{Gri94.1} (Fig.\,\ref{fig1}a). This structure persists down
to at least 40~mK, as shown by high-resolution neutron diffraction
measurements~\cite{Dee10.1}. The magnetic susceptibility $\chi(T)$ is
Curie-Weiss like above 100~K (Fig.\,\ref{fig1}b), with an effective moment close
to the full moment (2.54~$\mu_{\mathrm{B}}$/Ce) of the $J=5/2$ spin-orbit ground
state. A clear anomaly can be seen in $\chi(T)$ somewhat below $T_{\mathrm{N}}$
(Fig.\,\ref{fig1}c). The electrical resistivity $\rho(T)$ is typical of heavy
fermion compounds. $\Delta\rho$, the resistivity with the phonon-scattering
contribution subtracted, shows a $-\ln(T)$ behaviour due to incoherent Kondo
scattering at high temperatures. The maximum at about 20~K signals the onset of
Kondo screening (Fig.\,\ref{fig1}d). Also the Hall coefficient shows the typical
heavy fermion behaviour at high temperatures (Fig.\,\ref{fig1}e). Between room
temperature and 50~K it is well described by $R_{\mathrm{H}}(T) = R_0 +
R_{\mathrm{A}}(T)$, where $R_0$ represents a temperature independent normal Hall
coefficient and $R_{\mathrm{A}}(T)$ an intrinsic skew-scattering term. At low
temperatures, $R_{\mathrm{A}}$ becomes small and the measured Hall coefficient
is dominated by the normal Hall component (Supplementary Information).

The specific heat $C(T)$ reveals that, in addition to the phase transition at
$T_{\mathrm{N}}$ that is also seen in $\chi(T)$, there is a second phase
transition at $T_{\mathrm{Q}}$ (Fig\,\ref{fig1}f, ref.\,\onlinecite{Str06.2}).
The upper transition at $T_{\mathrm{Q}}$ has been tentatively attributed to
antiferro-quadrupolar order at the $8c$ site~\cite{Got09.1}, and the lower
transition at $T_{\mathrm{N}}$ to magnetic order~\cite{Mit10.1} -- presumably
antiferromagnetic order in analogy with Ce$_3$Pd$_{20}$Ge$_6$
(ref.\,\onlinecite{Doe00.1}). These assignments are consistent with the
$\Gamma_8$ quartet and the $\Gamma_7$ doublet of the crystalline electric field
split ground states of the Ce 4$f$ orbitals on the $8c$ and $4a$ sites,
respectively~\cite{Pas08.1,Dee10.1}. The temperature--field phase diagram is
shown in Fig.\,\ref{fig1}g. $T_{\mathrm{Q}}$ is initially enhanced by the
applied field but is eventually reduced at larger fields. According to
measurements on single crystals, magnetic field is able to completely suppress
$T_{\mathrm{Q}}$ (ref.\,\onlinecite{Mit10.1}), suggesting the presence of a
QCP at fields above 10~T. Within the ordered region $T_{\mathrm{Q}}(B) >
0$, $T_{\mathrm{N}}$ can be seen to decrease monotonically and vanish at about
0.9~T. This specifies a readily accessible QCP, thereby providing a rare
opportunity to study quantum phase transitions in cubic heavy fermion
materials. 

It follows from general symmetry considerations that antiferro-quadrupolar
ordering preserves the cubic symmetry of the lattice, as described in
Supplementary Information. In addition, such considerations as well as
microscopic calculations show that antiferro-quadrupolar order in the presence
of magnetic field induces dipolar order, thereby influencing the
antiferromagnetic correlations. The induced antiferromagnetic order, in turn,
implies a magnetic-field tuning of the antiferro-quadrupolar transition
temperature (Fig.~S5 of Supplementary Information), which is compatible with the
experimentally observed phase diagram (Fig.\,\ref{fig1}g). Through the magnetic
coupling between the $8c$ and $4a$ sites, the entire ordered region will contain
both magnetic and quadrupolar orders. Finally, in the absence of the competition
by the RKKY interactions,  the ground state multiplets at both sites will be
quenched by their Kondo couplings with the conduction electrons.

Studying the isothermal control parameter dependence of transport properties is
a well established means \cite{Pas04.1,Fri10.2} to probe the quantum critical
fluctuations near a QCP. Fig.\,\ref{fig2}a-c shows selected isotherms of the
Hall resistivity $\rho_{\mathrm{H}}$ as a function of the applied magnetic field
$B$. At the lowest temperatures $\rho_{\mathrm{H}}(B)$ shows two kinks
(Fig.\,\ref{fig2}a). One of these persists as a broadened feature at
temperatures above $T_{\mathrm{N}}$ (Fig.\,\ref{fig2}b,c). To quantify these
features we fit the data with crossover functions (Methods), shown as lines in
Fig.\,\ref{fig2}a-c. At temperatures above $T_{\mathrm{N}}$, this procedure
identifies the crossover field $B^{\ast}$, as well as the full width at half
maximum FWHM and the step height $\Delta A = |A_1 - A_2|$ of the crossover.
Below $T_{\mathrm{N}}$, the fitting characterizes in addition the crossover at
the N\'eel transition. The fitted quantities are shown in Fig.\,\ref{fig3}a-c
for the crossover at $B^{\ast}$ and in Fig.~S3 of Supplementary Information for
the crossover at $B_{\mathrm{N}}$. Broadened kinks in $\rho_{\mathrm{H}}(B)$
correspond to broadened steps in the differential Hall coefficient, defined as
the field derivative of the Hall resistivity $d\rho_{\mathrm{H}}/d B$. This is
shown in Fig.\,\ref{fig2}d for the crossover at $B^{\ast}$.

The features in the Hall resistivity have their counterparts in the longitudinal
and transverse magnetoresistance $\rho_{\mathrm{l}}(B)$ and
$\rho_{\mathrm{t}}(B)$ (Fig.\,\ref{fig2}e,f). The $B^{\ast}$ crossover appears
as a broadened step-like decrease of the resistivity with increasing field.
Below $T_{\mathrm{N}}$, the $B_{\mathrm{N}}$ crossover is seen as an increase of
the resistivity with field at small fields. The resistivity also contains a
component that increases more gradually with field. This is identified as a
background contribution (Methods). The resistivity data with the background and
the anomaly at $T_{\mathrm{N}}$ subtracted are shown in Fig.\,\ref{fig2}g,h and
are fitted by the same crossover functions that describe the differential Hall
coefficient. The fit parameters are also shown in Fig.\,\ref{fig3}a-c and
Fig.~S3 of Supplementary Information. 

Fig.\,\ref{fig3} demonstrates our key conclusions. $T^{\ast}(B)$ -- which is
equivalent to $B^{\ast}(T)$ -- defines a crossover scale that is distinct from
any phase transition line, except in the zero temperature limit where it merges
with $T_{\mathrm{N}}(B)$ at a common critical field of about 0.9~T
(Fig.\,\ref{fig3}a). The $T^{\ast}(B)$ scale exists both outside and within the
ordered part of the phase diagram delimited by the upper ordering temperature
$T_{\mathrm{Q}}(B)$. The FWHM of the crossover decreases with decreasing
temperature, extrapolating to zero in the zero temperature limit as evidenced by
the pure power-law behaviour of FWHM($T$) (Fig.\,\ref{fig3}b). At the same time
the step height $\Delta A$ remains finite in the zero temperature limit
(Fig.\,\ref{fig3}c). Because the Hall effect measures the response of the
electronic excitations near the Fermi surface, the crossover at nonzero
temperatures and the jump in the extrapolated zero-temperature limit are most
naturally interpreted in terms of a collapse of the heavy fermion Fermi surface
to a strongly reconstructed one. This implicates the $T^{\ast}$ line as
signifying a Kondo breakdown, which is tantamount to a localization of the $f$
electrons.

The observation of the collapsing Kondo breakdown scale implies new quantum
critical excitations which are neither of the Landau Fermi liquid nor of the spin
density wave QCP type \cite{Her76.1,Mil93.1}. Instead, electronic excitations
over the entire Fermi surface are expected to have a non-Fermi liquid form
\cite{Si01.1,Col01.1,Sen04.1}. In fact, the electrical resistivity is linear in
$T$ and the electronic specific heat coefficient $\Delta C(T)/T$ is logarithmic
in $T$ (refs.\,\onlinecite{Pas07.1,Str06.2}, Supplementary Information).
Both are defining characteristics of non-Fermi liquid behaviour which
appears also in other materials with Kondo breakdown QCPs. At magnetic fields
away from  $B^{\ast}(T=0)$, a Fermi liquid $T^2$ temperature dependence of the
electrical resistivity is recovered at low temperatures; measurements at several
magnetic fields (up to 5~T) suggest that the corresponding temperature scale
$T_{\rm FL}$ gradually decreases as $B$ approaches $B^{\ast}(T=0)$.

A collapsing Kondo breakdown scale has been observed in YbRh$_2$Si$_2$
(refs.\,\onlinecite{Pas04.1,Geg07.1,Fri10.2}). In that case the $T^{\ast}$ line
merges with the zero temperature boundary between paramagnetic and ordered
phases, thereby signaling the destruction of the Kondo effect and concomitant
reconstruction of the Fermi surface at the onset of  magnetic order
\cite{Si01.1,Col01.1,Sen04.1}. However, in Ce$_3$Pd$_{20}$Si$_6$ the $T^{\ast}$
line enters an ordered phase at finite temperature. We interpret this
distinction as due to the different dimensionality of the two compounds.

Spatial dimensionality modifies the degree of fluctuations, including that
of the quantum magnetism accociated with the $f$ moments. This is illustrated by
the two-parameter global phase diagram shown in Fig.\,\ref{fig4}. The horizontal
axis marks the strength $J_{\mathrm{K}}$ of the Kondo coupling between the local
$f$ moments and the conduction electrons. It controls the degree of quantum
fluctuations due to spin flip processes associated with the Kondo coupling. The
vertical axis $G$ describes the degree of quantum fluctuations within the local
moment component.

Going from the three-dimensional (3D) cubic limit to the decoupled 2D
limit amounts to moving upwards along the vertical axis. The tetragonal structure
of YbRh$_2$Si$_2$ suggests that it is close to the 2D limit, with enhanced $G$,
making it natural to have the ordered to paramagnetic phase boundary coinciding
with the Kondo collapse. The cubic structure of Ce$_3$Pd$_{20}$Si$_6$
implies a smaller $G$, placing it in the part of the global phase diagram where
Kondo collapsing occurs inside the ordered part of the phase diagram. Here, the
competition between the RKKY and Kondo couplings gives rise to a  $T^{\ast}$ line
which separates two ordered ground states. Note that, at finite temperature, the
$T^{\ast}$ line is distinct from the ordering transition lines. In the
zero-temperature limit, it separates a Kondo-screened order at $B>B^{\ast}$ and a
Kondo-destroyed order at $B<B^{\ast}$. Through the field-induced mixing of the
antiferro-quadrupolar and  antiferromagnetic orderings (Sec.\ E of Supplementary
Information), this corresponds to the region of the $G$--$J_{\mathrm{K}}$ phase
diagram in Fig.\,\ref{fig4} where a Kondo-destruction QCP (brown line) occurs
within the ordered part of the phase diagram, from a phase AF$_S$ to a phase
AF$_L$.

Our results provide a way to think about other quantum critical heavy fermion
metals, in line with recent theoretical considerations
\cite{Si06.1,Si10.1,Cus10.1,Col10.2}. CeIn$_3$ is another cubic system placing
it in a similar part of the vertical axis as Ce$_3$Pd$_{20}$Si$_6$. There is
some indication for the divergence of the effective quasiparticle mass inside
the ordered part of its phase diagram \cite{Seb09.1} making it instructive to
search for the $T^{\ast}$ scale in that system. The recently designed
CeIn$_3$/LaIn$_3$ superlattice \cite{Shi10.1} amounts to an elegant reduction of
the dimensionality towards the extreme 2D limit. Resistivity measurements have
already provided evidence for a reduced strength of magnetic ordering. It will
also be illuminating to explore the possibility of a Kondo breakdown. Finally,
there are materials which have 2D lattices that host geometrical frustration
such as the Shastry-Sutherland lattice in Ce$_2$Pt$_2$Pb
(ref.\,\onlinecite{Kim11.1}). It is possible that these materials have even
higer $G$ making them promising candidates to shed light on the upper part of
the global phase diagram.

Reconstruction of Fermi surface is also being extensively discussed in other
electronic materials, including cuprate superconductors \cite{Hel09.2}.
Typically, it is tied to antiferromagnetic ordering or other spontaneous
symmetry breaking transitions. Here, in Ce$_3$Pd$_{20}$Si$_6$, we find
Fermi-surface reconstructions both at the antiferromagnetic transition, the
$T_N$ line, and away from it, at the $T^{\ast}$ line. While the former is
smooth, the latter extrapolates to a jump of the Fermi surface in the
zero-temperature limit. Our results amount to a rare demonstration of
Fermi-surface reconstruction away from symmetry-breaking transitions. By
extension, our findings highlight the emergence of novel electronic excitations
through a mechanism other than spontaneous symmetry breaking, a notion that is
of considerable current interest in a variety of settings including topological
matters.

To summarize, we have observed an energy scale associated with the destruction
of the Kondo effect and the concominant $f$-electron localization in a cubic
heavy fermion compound. This not only extends the materials basis for this
effect to the 3D extreme but also unambiguously establishes that the origin of
the $T^{\ast}$ scale lies in robust many body correlations, as opposed to
materials specific band structure effects. Our findings suggest a materials
based global phase diagram for heavy fermion systems, which not only highlights
a rich variety of quantum critical points but also  indicates an underlying
universality. Given that quantum critical fluctuations represent an established
route towards unconventional superconductivity, the insight we have gained will
likely be important for the physics of high-temperature superconductors.

%METHODS !!!!!!!!!!!!!!!!!!!!!!!!!!!!!!!!!!!!!!!!!!!!!!!!!

{\noindent\large\bf Methods}

\noindent{\bf Synthesis and sample selection.} The polycrystalline samples were
prepared from high-purity elements (Ce 99.99\%, Pd 99.998\%, Si 99.9999\%) by
either ultra-high purity argon-arc or radio-frequency heating. Because of the
excellent stoichiometry of these polycrystals they are of higher quality than
the best available single crystals; this is evidenced by larger residual
resistance ratios, sharper phase transition anomalies and higher transition
temperatures in the polycrystals \cite{Pro09.1,Got09.1,Mit10.1}. The phase
transition temperature $T_{\mathrm{N}}(B)$ is isotropic in field down to the
lowest measured temperature \cite{Mit10.1}. We therefore chose these
polycrystals for our investigation.

\noindent{\bf Characterization.} The magnetotransport measurements were
performed by a standard 4-point ac technique in an Oxford dilution refrigerator
and, above 2~K, in a PPMS from Quantum Design. The magnetization measurements
were performed by a capacitive technique at low temperatures and in a SQUID
magnetometer of Cryogenic Ltd.\ above 2~K.

\noindent{\bf Data analysis.} The crossover in the magnetoresistance at
$B^{\ast}$ was fitted with the empirical crossover function
\begin{equation}
\rho(B) = A_2 - \frac{A_2-A_1}{1+(B/B^{\ast})^p}
\label{cross_ast}
\end{equation}
introduced in ref.\,\onlinecite{Pas04.1}, and the crossover at $T_{\mathrm{N}}$ with
the function
\begin{equation}
\rho(B) = A_2 - \frac{A_2-A_1}{1+e^{\frac{B-B_{\mathrm{N}}}{w}}}.
\label{cross_N}
\end{equation}
The latter function was chosen because it represents a single, symmetrically
broadened step of height $\Delta A$ and width $w$ at the finite field
$B_{\mathrm{N}}$, that describes the data very well. The Hall resistivity was
modeled with the integral over these fitting functions. 

As discussed in the main text, a smooth overall increase of $\rho$ with only
weak temperature dependence appears to be superimposed onto these two features.
Measuring the magnetoresistance in both the longitudinal (field parallel to
electrical current, Fig.\,\ref{fig2}e) and the transverse (field perpendicular
to electrical current, Fig.\,\ref{fig2}f) configuration helps us to identify
this latter as a background contribution due to normal magnetoresistance. It is
featureless at the QCP and should be eliminated for the analysis of quantum
criticality. We approximate the temperature dependent background by rescaling
the background functions of the lowest temperature isotherms (grey lines in
Fig.\,\ref{fig2}e,f) with the quadratic temperature dependence of the 15~T
resistivity data observed at the lowest temperatures. After the subtraction of
this background ($\rho_{\mathrm{l,back}}$ and $\rho_{\mathrm{t,back}}$) the
crossover-related magnetoresistance approaches zero at high fields. An exemplary
fit is shown in Fig.~S2 of Supplementary Information.

%REFERENCES !!!!!!!!!!!!!!!!!!!!!!!!!!!!!!!!!!!!!!!!!!!!!!!!!

\newpage

{\noindent\large\bf References}
%\bibliographystyle{nature}
%\bibliography{silke}
%\bibliography{/home/paschen/Silke/common/silke}
%\bibliography{silke_si}

\begin{thebibliography}{10}

\bibitem{Fri09.1}
Friedemann, S. \emph{et~al.}
\newblock {Detaching the antiferromagnetic quantum critical point from the
  Fermi-surface reconstruction in YbRh$_2$Si$_2$}.
\newblock {\em {Nature Phys.}}{ \bf {5}}, 465--469 ({2009}).

\bibitem{Tok09.1}
Tokiwa, Y., Gegenwart, P., Geibel, C. \& Steglich, F.
\newblock {Separation of energy scales in undoped YbRh$_2$Si$_2$ under
  hydrostatic pressure}.
\newblock {\em {J.\ Phys.\ Soc.\ Jpn.}}{ \bf 78}, 123708 ({2009}).

\bibitem{Cus10.1}
Custers, J. \emph{et~al.}
\newblock {Evidence for a non-Fermi-liquid phase in Ge-substituted
  YbRh$_2$Si$_2$}.
\newblock {\em {Phys.\ Rev.\ Lett.}}{ \bf 104}, 186402 (2010).

\bibitem{Sch10.1}
Schofield, A.~J.
\newblock {Quantum criticality and novel phases: Summary and outlook}.
\newblock {\em {Phys.\ Status Solidi B}}{ \bf 247}, 563--569 (2010).

\bibitem{Bro08.1}
Broun, D.~M.
\newblock {What lies beneath the dome?}
\newblock {\em {Nature Phys.}}{ \bf 4}, 170--172 (2008).

\bibitem{Loe94.1}
{v.\ L\"ohneysen}, H. \emph{et~al.}
\newblock {Non-Fermi-liquid behavior in a heavy-fermion alloy at a magnetic
  instability}.
\newblock {\em {Phys.\ Rev.\ Lett.}}{ \bf 72}, 3262--3265 (1994).

\bibitem{Mat98.1}
Mathur, N. \emph{et~al.}
\newblock {Magnetically mediated superconductivity in heavy fermion compounds}.
\newblock {\em {Nature}}{ \bf 394}, 39--43 (1998).

\bibitem{Geg07.1}
Gegenwart, P. \emph{et~al.}
\newblock {Multiple energy scales at a quantum critical point}.
\newblock {\em {Science}}{ \bf 315}, 969--971 (2007).

\bibitem{Par06.1}
Park, T. \emph{et~al.}
\newblock {Hidden magnetism and quantum criticality in the heavy fermion
  superconductor CeRhIn$_5$}.
\newblock {\em Nature}{ \bf 440}, 65--68 (2006).

\bibitem{Si06.1}
Si, Q.
\newblock {Global magnetic phase diagram and local quantum criticality in heavy
  fermion metals}.
\newblock {\em {Physica B}}{ \bf {378-380}}, 23--27 ({2006}).

\bibitem{Pas04.1}
Paschen, S. \emph{et~al.}
\newblock {Hall-effect evolution across a heavy-fermion quantum critical
  point}.
\newblock {\em {Nature}}{ \bf 432}, 881--885 (2004).

\bibitem{Fri10.2}
Friedemann, S. \emph{et~al.}
\newblock {Fermi-surface collapse and dynamical scaling near a quantum-critical
  point}.
\newblock {\em {PNAS}}{ \bf {107}}, 14547--14551 (2010).

\bibitem{Aro95.1}
Aronson, M. \emph{et~al.}
\newblock {Non-Fermi-liquid scaling of the magnetic response in
  UCu$_{5-x}$Pd$_x$ $(x=1,1.5)$}.
\newblock {\em {Phys.\ Rev.\ Lett.}}{ \bf {75}}, 725--728 (1995).

\bibitem{Sch00.1}
Schr\"{o}der, A. \emph{et~al.}
\newblock {Onset of antiferromagnetism in heavy-fermion metals}.
\newblock {\em {Nature}}{ \bf 407}, 351--355 (2000).

\bibitem{Gri94.1}
Gribanov, A.~V., Seropegin, Y.~D. \& Bodak, O.~I.
\newblock {Crystal structure of the compounds Ce$_3$Pd$_{20}$Ge$_6$ and
  Ce$_3$Pd$_{20}$Si$_6$}.
\newblock {\em {J.\ Alloys Compd.}}{ \bf 204}, L9--L11 (1994).

\bibitem{Dee10.1}
Deen, P.~P. \emph{et~al.}
\newblock {Quantum fluctuations and the magnetic ground state of
  Ce$_3$Pd$_{20}$Si$_6$}.
\newblock {\em {Phys.\ Rev.\ B}}{ \bf 81}, 064427 (2010).

\bibitem{Str06.2}
Strydom, A.~M., Pikul, A., Steglich, F. \& Paschen, S.
\newblock {Possible field-induced quantum criticality in
  Ce$_3$Pd$_{20}$Si$_6$}.
\newblock {\em {J.\ Phys.: Conf.\ Series}}{ \bf 51}, 239--242 (2006).

\bibitem{Got09.1}
Goto, T. \emph{et~al.}
\newblock {Quadrupole ordering in clathrate compound
  Ce$_{3}$Pd$_{20}$Si$_{6}$}.
\newblock {\em {J.\ Phys.\ Soc.\ Jpn.}}{ \bf 78}, 024716 (2009).

\bibitem{Mit10.1}
Mitamura, H. \emph{et~al.}
\newblock {Low temperature magnetic properties of Ce$_3$Pd$_{20}$Si$_6$}.
\newblock {\em {J.\ Phys.\ Soc.\ Jpn.}}{ \bf 79}, 074712 (2010).

\bibitem{Doe00.1}
D\"onni, A. \emph{et~al.}
\newblock {Low-temperature antiferromagnetic moments at the 4a site in
  Ce$_3$Pd$_{20}$Ge$_6$}.
\newblock {\em {J.\ Phys.: Condens.\ Matter}}{ \bf 12}, 9441--9451 (2000).

\bibitem{Pas08.1}
Paschen, S. \emph{et~al.}
\newblock {First neutron measurements on Ce$_3$Pd$_{20}$Si$_6$}.
\newblock {\em {Physica B}}{ \bf 403}, 1306--1308 (2008).

\bibitem{Her76.1}
Hertz, J.
\newblock {Quantum critical phenomena}.
\newblock {\em {Phys.\ Rev.\ B}}{ \bf 14}, 1165--1184 (1976).

\bibitem{Mil93.1}
Millis, A.~J.
\newblock {Effect of a nonzero temperature on quantum critical points in
  itinerant fermion systems}.
\newblock {\em {Phys.\ Rev.\ B}}{ \bf 48}, 7183--7196 (1993).

\bibitem{Si01.1}
Si, Q., Rabello, S., Ingersent, K. \& Smith, J.
\newblock {Locally critical quantum phase transitions in strongly correlated
  metals}.
\newblock {\em Nature}{ \bf 413}, 804 (2001).

\bibitem{Col01.1}
Coleman, P., P\'epin, C., Si, Q. \& Ramazashvili, R.
\newblock {How do Fermi liquids get heavy and die?}
\newblock {\em {J.\ Phys.: Condens.\ Matter}}{ \bf 13}, R723 (2001).

\bibitem{Sen04.1}
Senthil, T., Vojta, M. \& Sachdev, S.
\newblock {Weak magnetism and non-Fermi liquids near heavy-fermion critical
  points}.
\newblock {\em {Phys.\ Rev.\ B}}{ \bf {69}}, 035111 ({2004}).

\bibitem{Pas07.1}
Paschen, S. \emph{et~al.}
\newblock {Quantum critical behaviour in Ce$_3$Pd$_{20}$Si$_6$?}
\newblock {\em {J.\ Magn.\ Magn.\ Mater.}}{ \bf {316}}, 90--92 (2007).

\bibitem{Si10.1}
Si, Q.
\newblock {Global magnetic phase diagram and local quantum criticality in heavy
  fermion metals}.
\newblock {\em {Phys.\ Status Solidi}}{ \bf {247}}, 476--484 ({2010}).

\bibitem{Col10.2}
Coleman, P. \& Nevidomskyy, A.
\newblock {Frustration and the Kondo effect in heavy fermion materials}.
\newblock {\em {J.\ Low Temp.\ Phys.}}{ \bf 161}, 182--202 (2010).

\bibitem{Seb09.1}
Sebastian, S.~E. \emph{et~al.}
\newblock {Heavy holes as a precursor to superconductivity in antiferromagnetic
  CeIn$_3$}.
\newblock {\em {PNAS}}{ \bf {106}}, 7741--7744 (2009).

\bibitem{Shi10.1}
Shishido, H. \emph{et~al.}
\newblock {Tuning the dimensionality of the heavy fermion compound CeIn$_3$}.
\newblock {\em {Science}}{ \bf 327}, 980--983 (2010).

\bibitem{Kim11.1}
Kim, M.~S. \& Aronson, M.~C.
\newblock {Heavy fermion compounds on the geometrically frustrated
  Shastry-Sutherland lattice}.
\newblock {\em {J.\ Phys.: Condens.\ Matter}}{ \bf 23}, 164204 (2011).

\bibitem{Hel09.2}
Helm, T. \emph{et~al.}
\newblock {Evolution of the Fermi surface of the electron-doped
  high-temperature superconductor
  ${\mathrm{Nd}}_{2-x}{\mathrm{Ce}}_{x}{\mathrm{CuO}}_{4}$ revealed by
  Shubnikov--de Haas oscillations}.
\newblock {\em {Phys.\ Rev.\ Lett.}}{ \bf 103}, 157002 (2009).

\bibitem{Pro09.1}
Prokofiev, A. \emph{et~al.}
\newblock {Crystal growth and composition-property relationship of
  Ce$_3$Pd$_{20}$Si$_6$ single crystals}.
\newblock {\em {Phys.\ Rev.\ B}}{ \bf 80}, 235107 (2009).

\end{thebibliography}

\vspace{0.5cm}

{\noindent\large\bf Acknowledgements}\\
The authors wish to thank  S. Kirchner for  useful discussions. The work was
funded by the European Research Council under the European Community's Seventh
Framework Programme (FP7/2007-2013)/ERC grant agreement no. 227378 and by the
Austrian Science Foundation (project P19458-N16). A.St.\ thanks the SA-NRF
(2072956) and the URC of the University of Johannesburg for financial
assistance. R.Y.\ and Q.S.\ acknowledge the support of  NSF Grant No.\
DMR-1006985 and the Robert A.\ Welch Foundation Grant No.\ C-1411.

\vspace{0.5cm}

{\noindent\large\bf Author contributions}\\
S.P.\ initiated the study. S.P.\ and Q.S.\ designed the research. A.St.\ and
A.P.\ synthesized and characterized the material. J.C., K.L., M.M., and H.W.\
performed magnetotransport measurements, A.Si.\ and Y.S.\ magnetization
measurements. T.S.\ led the low-temperature magnetization investigation. K.L.,
H.W., A.Si., and S.P.\ analyzed the data. R.Y.\ and Q.S.\ set up the theoretical
framework and performed the calculations. S.P., Q.S., and R.Y.\ prepared the
manuscript. All authors contributed to the discussion.

\vspace{0.5cm}

{\noindent\large\bf Additional Information}\\
The authors declare that they have no competing financial interests.
Supplementary information accompanies this paper on
www.nature.com/naturematerials. Reprints and permission information is available
online at http://www.nature.com/reprints. Correspondence and requests for
materials should be addressed to S.P.

%FIGURE LEGENDS !!!!!!!!!!!!!!!!!!!!!!!!!!!!!!!!!!!!!!!!!!!!!!!!!

\newpage
\begin{figure}[h!]
\caption{\label{fig1} Characteristics of the heavy fermion compound
Ce$_3$Pd$_{20}$Si$_6$. {\bf a}, Cubic crystal structure. {\bf b}, Inverse volume
susceptibility in SI units $1/\chi^{\mathrm{SI}}$ in the linear response regime
vs temperature $T$. A Curie-Weiss fit at high temperatures yields an effective
moment $\mu = 2.35 \mu_{\mathrm{B}}$ per Ce atom and a paramagnetic Weiss
temperature $\Theta = -3$~K. {\bf c}, A maximum in $\chi^{\mathrm{SI}}(T)$
somewhat below $T_{\mathrm{N}}$ clearly reveals the N\'eel transition. Only a
very weak feature can be discerned at the putative antiferro-quadrupolar
transition at $T_{\mathrm{Q}}$. {\bf d}, Temperature-dependent electrical
resistivity $\rho(T)$ and $\Delta\rho(T) = \rho(T) - \rho_{\mathrm{ph}}(T)$,
where the contribution due to phonon scattering $\rho_{\mathrm{ph}}(T)$ is
determined from the non-$f$ reference compound La$_3$Pd$_{20}$Si$_6$ as in
ref.\,\onlinecite{Str06.2}. {\bf e}, Temperature dependent Hall coefficient in
the linear response regime $R_{\mathrm{H}}(T)$, together with a fit according to
the anomalous Hall effect model described in Supplementary Information, and its
extrapolation to lower temperatures (dashed line). The normal Hall coefficient
$R_0$, assumed as temperature independent in this model, is shown as grey line.
{\bf f}, Electronic contribution to the specific heat $\Delta C$ vs $T$ in an
applied magnetic field of 0.5~T (from ref.\,\onlinecite{Str06.2}). The two
anomalies at $T_{\mathrm{N}}$ and $T_{\mathrm{Q}}$ are clear signatures of
second-order phase transitions. It is expected that their ordering wavevectors
are different. As in ref.\,\onlinecite{Pas07.1} the transition temperatures are
estimated by entropy balance constructions. {\bf g}, Temperature-field phase
diagram with the transition temperatures $T_{\mathrm{N}}$ and $T_{\mathrm{Q}}$,
determined from transport and specific heat measurements, respectively.}
\end{figure}

\begin{figure}[h!]
\caption{\label{fig2} Magnetotransport across the quantum critical point of
Ce$_3$Pd$_{20}$Si$_6$. {\bf a}, Hall resistivity $\rho_{\mathrm{H}}$ vs applied
magnetic field $B$ at different temperatures below $T_{\mathrm{N}}$. The solid
lines represent fits of a crossover function (see Methods) to the data. {\bf b,
c}, Corresponding plots for intermediate and high temperatures. The crossover
fields $B_{\mathrm{N}}$ and $B^{\ast}$ of the fits are indicated in {\bf a}-{\bf
c}. {\bf d}, Field derivative of the fits of {\bf a}-{\bf c},
$d\rho_{\mathrm{H}}/dB$, normalized to the step height $\Delta A$, vs normalized
field $B/B^{\ast}$. The extrapolated zero-temperature form, a sharp step, is
shown as grey line. {\bf e}, Longitudinal magnetoresistance $\rho_{\mathrm{l}}$
vs $B$ at different temperatures (80, 100, 125, 150, 175, 200, 250, 301, 350,
402, 500, 602, 700, 900~mK, 1.1, 1.3, 1.5, 1.7, 1.9, 2.9, 5.0~K). {\bf f},
Corresponding plot for transverse magnetoresistance $\rho_{\mathrm{t}}$ (89,
193, 300, 366, 486, 632, 743, 963~mK, 1.1, 1.3, 1.5, 1.7, 2.0, 3.0, 5.0~K). The
grey curves in {\bf e} and {\bf f} represent the background (see Methods). {\bf
g}, Crossover component at $B^{\ast}$ (see Methods) of $\rho_{\mathrm{l}}(B)$,
normalized to the zero-field resistivity at the respective temperature, vs
normalized field $B/B^{\ast}$. The data points are shown as dots, the fits as
full lines. The grey line again represents the extrapolated zero-temperature
form. {\bf h}, Corresponding plot for $\rho_{\mathrm{t}}(B)$. }
\end{figure}

\begin{figure}[h!]
\caption{\label{fig3} Characteristics of the Fermi surface collapse in
Ce$_3$Pd$_{20}$Si$_6$. {\bf a}, Temperature $T^{\ast}(B)$ of the crossover at
$B^{\ast}(T)$, determined from the fits of $\rho_{\mathrm{H}}(B)$,
$\rho_{\mathrm{l}}(B)$, and $\rho_{\mathrm{t}}(B)$ in Fig.\,\ref{fig2}, plotted
in the temperature-field phase diagram of Fig.\,\ref{fig1}g. {\bf b},
Temperature dependence of the full width at half maximum, FWHM($T$), of the
crossover. {\bf c}, Temperature dependence of the step height, $\Delta A(T)$, of
the crossover. The crosses refer to Hall resistivity data corrected for the
anomalous Hall effect (Supplementary Information).}
\end{figure}

\newpage

\begin{figure}[ht!]
\caption{\label{fig4} Materials-based global phase diagram for heavy fermion
compounds near antiferromagnetic instabilities. Magnetic frustration parameter
$G$ (left) vs Kondo coupling $J_{\mathrm{K}}$ at $T = 0$. Lines of quantum
critical points separate antiferromagnetic (AF) from paramagnetic (P) regions
(thick red line), and regions of small (S) and large (L) Fermi surface (brown
line). The latter line represents quantum critical points accompanied by a Kondo
breakdown. Dimensionality (right) helps to calibrate the placement of selected
materials (CPS: Ce$_3$Pd$_{20}$Si$_6$, CeIn$_3$, a CeIn$_3$/LaIn$_3$
superlattice, YRS: YbRh$_2$Si$_2$, CCA: CeCu$_{6-x}$Au$_x$, CPP: Ce$_2$Pt$_2$Pb,
marked on the $G$-axis by the ticks) along the vertical axis. The present work
allows to elucidate the three-dimensional part of the phase diagram.}
\end{figure}

% FIGURES !!!!!!!!!!!!!!!!!!!!!!!!!!!!!!!!!!!!!!!!!!!!!!!!!!!!!!

\newpage

\begin{figure}[t!]
\centerline{\includegraphics[width=140mm]{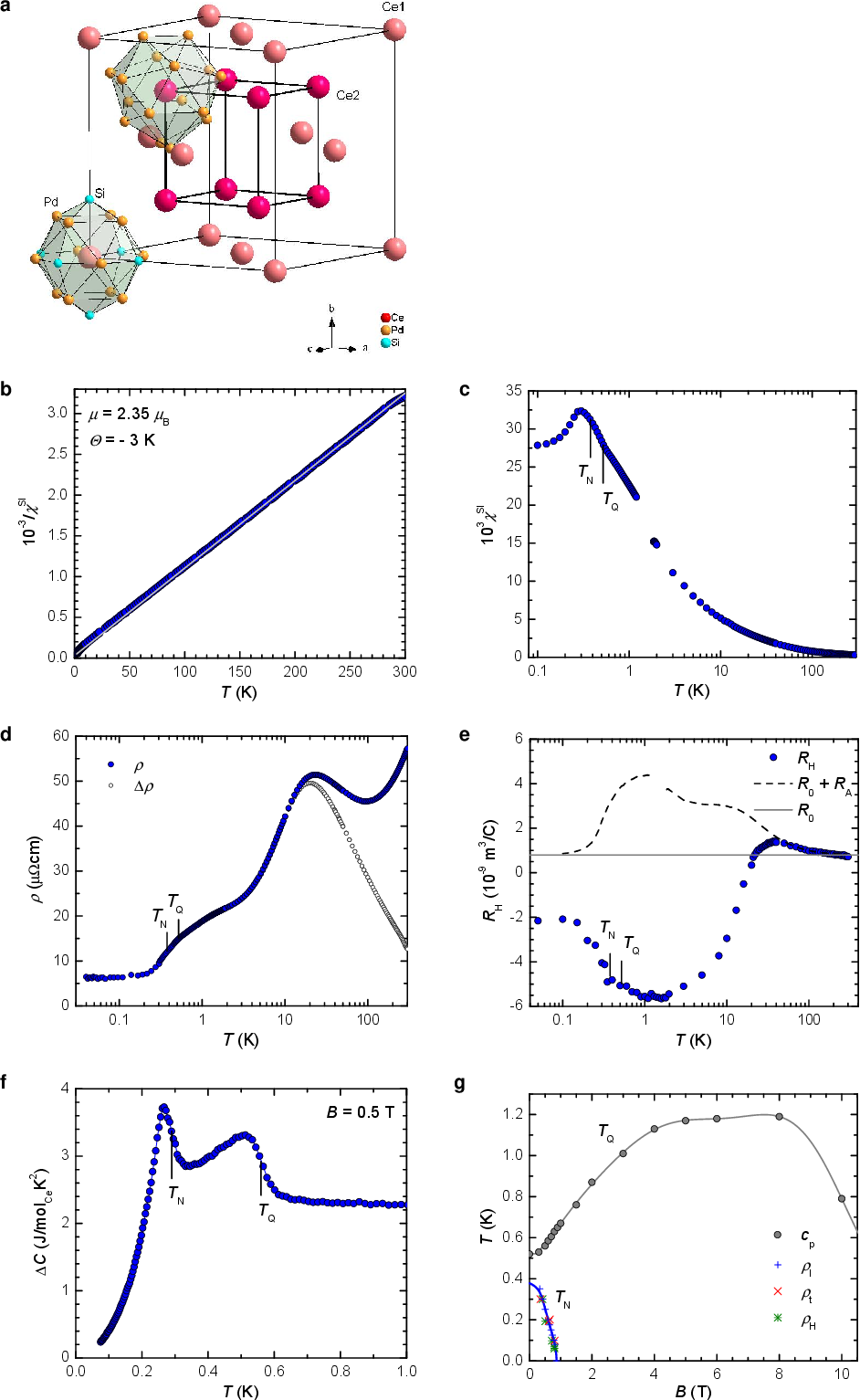}}
{\Large Figure 1}
\end{figure}

\newpage

\begin{figure}[t!]
\centerline{\includegraphics[width=150mm]{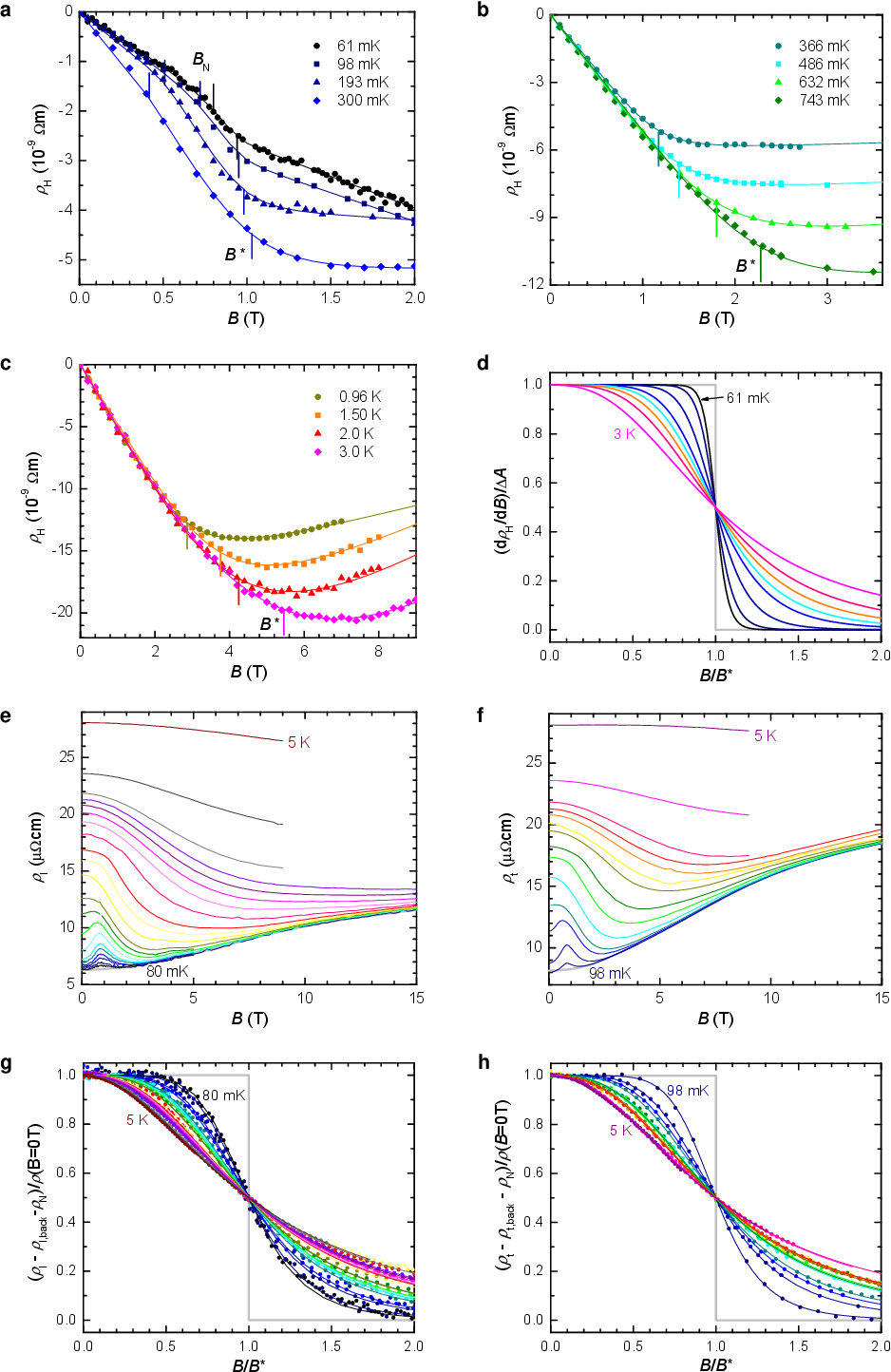}}
{\Large Figure 2}
\end{figure}

\newpage

\begin{figure}[t!]
\centerline{\includegraphics[width=100mm]{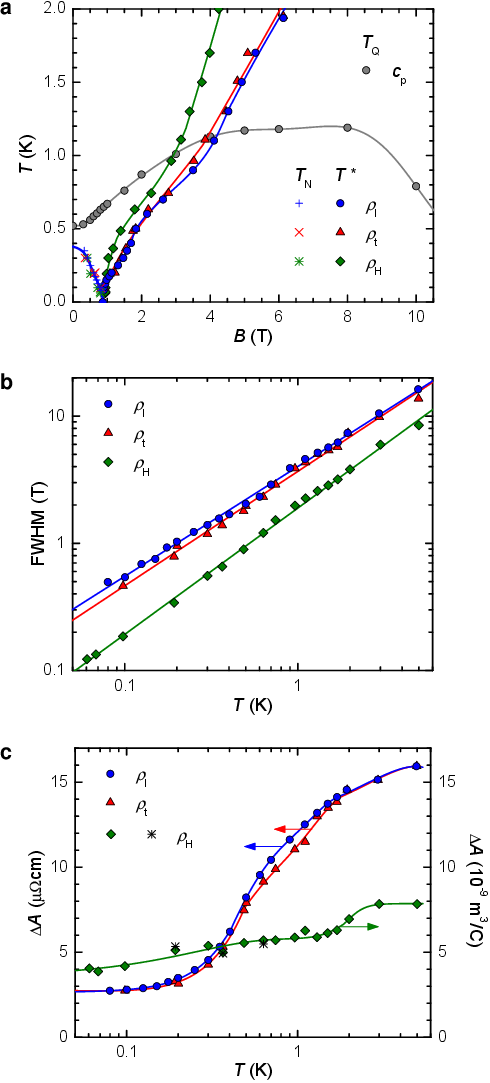}}
{\Large Figure 3}
\end{figure}

\begin{figure}[t!]
\centerline{\includegraphics[width=140mm]{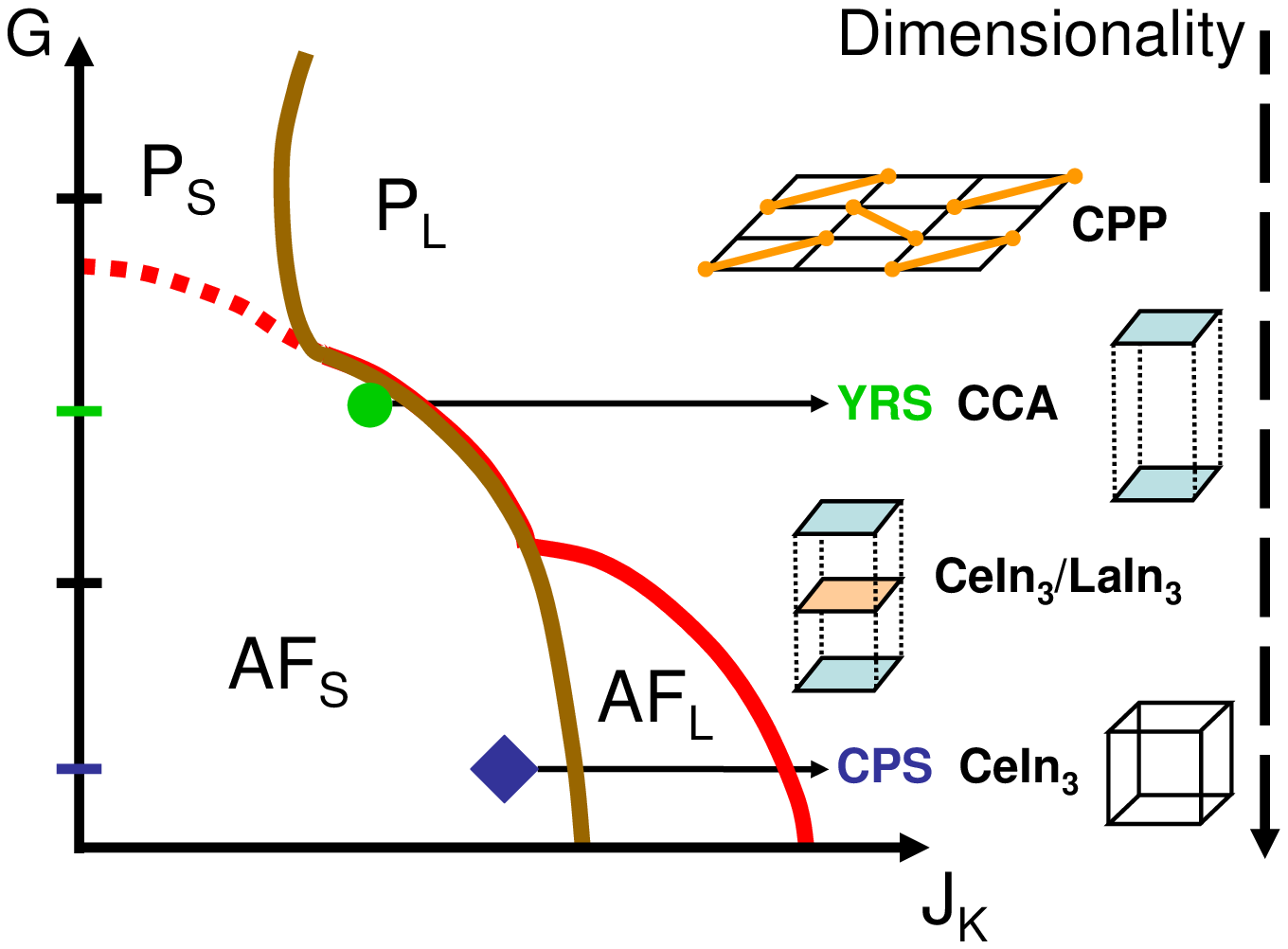}}
{\Large Figure 4}
\vspace{8cm}
\end{figure}

\end{document}

\hyphenation{va-ni-sh-ing}

\begin{center}
{\large\bf Destruction of Kondo effect in cubic heavy fermion compound
Ce$_3$Pd$_{20}$Si$_6$: Supplementary Information}
\\[0.5cm]

J.\ Custers$^1$, K.-A.\ Lorenzer$^1$, M.\ M\"uller$^1$, A.\ Prokofiev$^1$, A.\
Sidorenko$^1$, H.\ Winkler$^1$, A.\ M.\ Strydom$^2$, Y.\ Shimura$^3$, T.\
Sakakibara$^3$, R.\ Yu$^4$, Q.\ Si$^4$, and S.\ Paschen$^{1\ast}$\\

$^1$ Institute of Solid State Physics, Vienna University of Technology, 1040
Vienna, Austria\\

$^2$ Physics Department, University of Johannesburg, Auckland Park 2006, South
Africa\\

$^3$ Institute for Solid State Physics, University of Tokyo, Kashiwa 277-8581,
Japan\\

$^4$ Department of Physics and Astronomy, Rice University, Houston, TX 77005,
USA\\

$^{\ast}$ e-mail: paschen@ifp.tuwien.ac.at

\end{center}

{\noindent\bf A. Non-Fermi liquid properties}

Non-Fermi liquid behaviour is observed in Ce$_3$Pd$_{20}$Si$_6$, as has been
reported previously \cite{Pas07,Str06}. At a field of 1~T, close to the
critical field of 0.9~T, the low-temperature electrical resistivity is linear in
$T$ (Fig.\,S1a, ref.\,\onlinecite{Pas07}) and the electronic specific heat
coefficient $\Delta C(T)/T$ is logarithmic in $T$ (Fig.\,S1b,
ref.\,\onlinecite{Str06}). This is to be contrasted with the behaviour
expected at a 3D spin density wave QCP, namely a $T^{3/2}$ temperature
dependence for the electrical resistivity and $A - B\sqrt T$ for $\Delta
C(T)/T$. At fields further away from the critical field, Landau Fermi liquid
behaviour is recovered at lowest temperatures as clearly seen in the form of a
$T^2$ dependence of the electrical resistivity. At higher temperatures, however,
non-Fermi liquid characetristics are observed for a broader field range
indicating that a fan of quantum critical behaviour emanates from the QCP with
increasing temperature and extends even to above $T_{\mathrm{Q}}$. Note the very
large value of the electronic specific heat coefficient that reaches more than
7~J/mol$_{\mathrm{Ce}}$K$^2$ in the non-Fermi liquid region.

{\noindent\bf B. Fitting the crossover data}

The magnetoresistivity with crossovers at $B^{\ast}$ and $B_{\mathrm{N}}$ was
analyzed with the sum of two empirical fitting functions (equations
(1) and (2), Methods). Figure~S2 shows a
typical result for the longitudinal magnetoresistivity, with the background
contribution subtracted. The grey line is the sum of the two functions. It
describes the data very well. Above the N\'eel temperature, only the crossover
at $B^{\ast}$ remains and thus the data were fit to equation (1).

{\noindent\bf C. Crossover at the N\'eel transition}

The fit parameters characterizing the crossover at $B^{\ast}$ were given in
Fig.~3 of the main part. In Fig.\,S3 we show the corresponding
quantities for the crossover at the N\'eel transition. Both the step height
$\Delta A_{\mathrm{N}}$ and the crossover width $w_{\mathrm{N}}$ increase with
decreasing temperature. This is in contrast to the crossover at $B^{\ast}$ which
becomes infinitely sharp in the extrapolated zero temperature limit.

{\noindent\bf D. Anomalous Hall effect}

As is typical for heavy fermion metals the temperature dependence of the Hall
coefficient $R_{\mathrm{H}}$ at high temperatures is dominated by an anomalous
Hall component due to intrinsic skew-scattering, $R_{\mathrm{A}}(T) = \gamma
\rho_{\mathrm{mag}}(T) \chi(T)$, where $\rho_{\mathrm{mag}}(T) = \rho(T) -
\rho_{\mathrm{ph}}(T) - \rho_0$ is the intrinsic magnetic contribution to the
electrical resistivity $\rho(T)$, $\rho_{\mathrm{ph}}(T)$ is the phonon
contribution determined via the non-magnetic reference compound
La$_3$Pd$_{20}$Si$_6$, $\rho_0$ is the residual resistivity, $\chi(T)$ is the
magnetic susceptibility, and $\gamma$ is a positive constant~\cite{Fer87}. This
was shown in Fig.~1e of the main part where, in the incoherent Kondo
regime above 50~K, the measured $R_{\mathrm{H}}(T)$ is well described by
the sum of a temperature independent normal Hall coefficient $R_0$ and
$R_{\mathrm{A}}(T)$. The constant $\gamma = (1.29 \pm 0.03)$~m$^2$/Vs was
determined by plotting $R_{\mathrm{H}}$ vs $\rho_{\mathrm{mag}} \chi$, with
temperature as implicit parameter, which varies approximately linearly between
room temperature and 50~K (not shown). The $R_0 + R_{\mathrm{A}}$ curve below
50~K was determined by assuming, as in refs.\,\onlinecite{Pas04,Fri10}, that
$\gamma$ remains temperature independent down to the lowest temperatures. As
seen in Fig.~1e of the main part, $R_{\mathrm{A}}(T)$ remains positive
and becomes negligibly small at the lowest temperatures. The measured Hall
coefficient, on the other hand, becomes negative below about 20~K and saturates
at the lowest temperatures. It must, therefore, be dominated by a temperature
dependent normal Hall component.

In addition to the intrinsic skew-scattering term, there is an extrinsic
contribution due to skew-scattering defects that might become relevant at the
lowest temperatures. In order to estimate its magnitude we assume that, as an
upper boundary, the entire difference of the residual resistivities of
Ce$_3$Pd$_{20}$Si$_6$ and La$_3$Pd$_{20}$Si$_6$, about 10\% of $\rho_0$, is
skew-scattering active. The extrinsic anomalous Hall effect then amounts to 10\%
of $R_{\mathrm{H}}$ at the lowest accessed temperature.

Finally, we clarify whether the magnetic field dependence of the measured Hall
resistivity $\rho_{\mathrm{H}}(B)$ is perturbed by an anomalous contribution
$\rho_{\mathrm{H}}^{\mathrm{a}}(B) = \gamma \rho_{\mathrm{skew}}(B) \mu_0 M(B)$.
$\rho_{\mathrm{skew}}(B)$ contains the temperature and field dependent
background contribution to the longitudinal resistivity $\rho_{\mathrm{l,back}}$
(see Methods) and the above discussed 10\% of $\rho_0$. In Fig.\,S4 the normal
component to the Hall resistivity $\rho_{\mathrm{H}}^0(B) = \rho_{\mathrm{H}}(B)
- \rho_{\mathrm{H}}^{\mathrm{a}}$ is shown at three selected temperatures,
together with the $\rho_{\mathrm{H}}$ data and their fits.
$\rho_{\mathrm{H}}^{\mathrm{a}}$ reaches at maximum 20\% of $\rho_{\mathrm{H}}$.
Interestingly, $\rho_{\mathrm{H}}^0(B)$ is very well described by crossover
functions that differ from the fits of $\rho_{\mathrm{H}}(B)$ only by slightly
different crossover heights $\Delta A$. These are included as stars in
Fig.~3c of the main part and Fig.\,S3a. Thus, taking the anomalous
Hall effect into account has no significant effect on the outcome of our
analyses.

{\noindent\bf E. Antiferromagnetic order induced by antiferro-quadrupolar order:
microscopic and symmetry-based analyses}

In Ce$_3$Pd$_{20}$Si$_6$, the Ce$^{3+}$ ion has one localized $f$ electron
with angular momentum $J=5/2$. In cubic environment, the six-fold degenerate
ground-state multiplet is further split into a $\Gamma_8$ quartet and a
$\Gamma_7$ doublet. We will take the experimentally suggested~\cite{Deen10}
ground state of the Ce ion, which is a $\Gamma_7$ doublet for the $4a$ site and a
$\Gamma_8$ quartet for the $8c$ site. Building upon the suggestion of ultrasound
measurements~\cite{Goto09}, we associate the high-temperature transition at
$T_{\mathrm{Q}}$ with antiferro-quadrupolar (AFQ) order of the $\Gamma_8$ subsystem of Ce
ions on the $8c$ site.

In this section, we  show that, in the presence of a magnetic field, the AFQ 
order is expected to induce antiferromagnetic (AFM) order. This is seen through
both a microscopic analysis of an effective pseudospin model, as well as a
symmetry-based analysis of a Ginzburg-Landau functional. We also demonstrate
that the AFQ order preserves the cubic structural symmetry, and that the
experimentally observed evolution of the AFQ order as a function of the magnetic
field is compatible with both the microscopic model calculation and
symmetry-based considerations.

{\noindent\it Effective pseudospin model}

We will concentrate  on the Ce ions  of the $8c$ site, which form a simple cubic
sublattice. The $\Gamma_8$ quartet can be described  through two pseudospin
operators~\cite{Shiina97}, $\boldsymbol{\sigma}$ and $\boldsymbol{\tau}$, which
respectively act on the Kramers and non-Kramers doublet states. We can then
express the RKKY  coupling in terms of these pseudospin operators in a
Kugel-Khomskii way~\cite{Ohkawa1,Ohkawa2}:
\begin{eqnarray}\label{Eq:ST01}
 \mathcal{H} &=& J\sum_{\langle i,j\rangle} \left[ \boldsymbol{\tau}_i\cdot
\boldsymbol{\tau}_j + \boldsymbol{\sigma}_i\cdot \boldsymbol{\sigma}_j
+4(\boldsymbol{\tau}_i\cdot \boldsymbol{\tau}_j)(\boldsymbol{\sigma}_i\cdot
\boldsymbol{\sigma}_j)\right]+\mathcal{H}_Z,
\label{K-K}
\end{eqnarray}
where $J$ is the nearest neighbor coupling strength,  and $\mathcal{H}_Z$ is the
Zeeman term  associated with the applied magnetic field
$\mathbf{H}=(H_x,H_y,H_z)$.

The $\Gamma_8$ system  comprises three dipoles, five quadrupoles and seven
octupoles, all of which are  irreducible representations of the $O_h$ group  of
the cubic lattice. We denote  them by $\boldsymbol\tau^\prime$,
$\boldsymbol\mu$, $\boldsymbol\sigma$, $\boldsymbol\eta$, $\boldsymbol\zeta$,
and $\xi$ (defined in terms of  $\boldsymbol{\tau}$ and $\boldsymbol{\sigma}$,
Table~\ref{TableSTI}), and rewrite equation~(\ref{Eq:ST01}) as
\begin{eqnarray} \label{Eq:ST03}
 \mathcal{H} &=& \sum_{\langle i,j\rangle} \left[
J_\tau\boldsymbol{\tau^\prime}_i\cdot \boldsymbol{\tau^\prime}_j +
J_\mu\boldsymbol{\mu}_i\cdot \boldsymbol{\mu}_j +
J_\sigma\boldsymbol{\sigma}_i\cdot \boldsymbol{\sigma}_j  \right.
+ J_\eta\boldsymbol{\eta}_i\cdot \boldsymbol{\eta}_j +
\left.J_\zeta\boldsymbol{\zeta}_i\cdot \boldsymbol{\zeta}_j + J_\xi\xi_i\cdot
\xi_j\right]+\mathcal{H}_Z.
\end{eqnarray}
Here, the first two terms refer to quadrupole-quadrupole interactions; the third
and fourth terms  are couplings  among both  the dipoles  and  octupoles; the
next two terms  are purely among the octupoles. By allowing the couplings to
take different values in equation~(\ref{Eq:ST03}),  we have extended 
equation~(\ref{Eq:ST01})  to a more general (and realistic)  microscopic
Hamiltonian that is compatible with the cubic lattice symmetry.  The different
coupling values also lift the degeneracy among the different  types of
multipoles. In order to stabilize the AFQ order, we take the first two couplings
to be positive; through equation~(\ref{K-K}) the remaining couplings will also
be taken as positive. The ultrasound experiment~\cite{Goto09} has been
interpreted in terms of an AFQ phase with an ordering wavevector
$(\pi,\pi,\pi)$, although it is also compatible with a $(q,q,q)$ AFQ order (and
its symmetry equivalents) for any nonzero $q$. The Zeeman term $\mathcal{H}_Z$
reads $\mathcal{H}_Z = -\frac{7}{3}g\mu_B\sum_i
(\boldsymbol{\sigma}_i+\frac{4}{7}\boldsymbol\eta_i)\cdot \mathbf{H}$, where
$g=6/7$ is the appropriate gyromagnetic factor.

The quadrupolar moments may couple to elastic strain via~\cite{Goto09}
$\mathcal{H}_{QS}=-\sum_{i,\delta} g_\delta O_\delta(i) \varepsilon_\delta$,
where $\delta=x^2-y^2,2z^2-x^2-y^2,xy,xz,yz$, $\varepsilon_\delta$ refers to the
elastic strain of the $\delta$'s component, and $g_\delta$ is the corresponding
coupling constant. Through this coupling to elastic distortion, quadrupolar
order may spontaneously break the cubic lattice symmetry. This is the case in
the Ce$_3$Pd$_{20}$Ge$_6$ system~\cite{CPG1,CPG2}, where the ferroquadrupolar order
reduces the lattice symmetry from cubic to tetragonal.  However, in our model
for Ce$_3$Pd$_{20}$Si$_6$ the $(\pi,\pi,\pi)$ AFQ order has a zero net quadrupolar moment, which
could not directly couple to elastic distortion. In other words, the AFQ order
preserves the cubic symmetry. We will therefore retain the cubic lattice
symmetry in the remainder of the analysis.

{\noindent\it Microscopic studies of the AFQ and AFM orders of the effective
pseudospin model}

In order to address the coupling between the AFQ and AFM orders in the presence
of a magnetic field, as well as the field evolution of $T_{\mathrm{Q}}$, we solve the model
Hamiltonian of equation~(\ref{Eq:ST03}) at the mean-field level. For each pair
of pseudospin operators $\boldsymbol\phi_i\cdot\boldsymbol\phi_j$ (where
$\boldsymbol\phi=\boldsymbol\tau^\prime$, $\boldsymbol\mu$, $\boldsymbol\sigma$,
$\boldsymbol\eta$, $\boldsymbol\zeta$, $\xi$) we perform a mean-field
decomposition $\boldsymbol\phi_i\cdot\boldsymbol\phi_j\approx
\langle\boldsymbol\phi_i\rangle\cdot\boldsymbol\phi_j +
\boldsymbol\phi_i\cdot\langle\boldsymbol\phi_j\rangle -
\langle\boldsymbol\phi_i\rangle\cdot\langle\boldsymbol\phi_j\rangle$. We assume
a two-sublattice AFQ order with ordering vector $(\pi,\pi,\pi)$ [or
$(q, q, q)$], introduce the uniform and staggered components of the
mean-field parameters as $\boldsymbol\phi_f=(\langle\boldsymbol\phi_A\rangle +
\langle\boldsymbol\phi_B\rangle)/2$ and
$\boldsymbol\phi_a=(\langle\boldsymbol\phi_A\rangle -
\langle\boldsymbol\phi_B\rangle)/2$ for the sublattices A and B, and minimize
the free energy.

As suggested by ultrasound measurements~\cite{Goto09}, the coupling strength
between $\Gamma_5^+$ quadrupolar moments is stronger than the one between
$\Gamma_3^+$ moments. To reflect this in the model, we set $J_\tau=1$, and take
$J_\mu>J_\tau$. Since we are mainly interested in the ordering of quadrupolar
and dipolar moments, we  reduce the coupling strengths of the pure octupolar
moments  by hand. Without loss of generality, we will work with the following
parameters: $J_\tau=1$, $J_\mu=1.1$, $J_\sigma=J_\eta=1$, and
$J_\zeta=J_\xi=0.75$.

We study the finite-temperature magnetic phase diagram in the presence of an
external magnetic field. The direction of the magnetic field can be expressed by
two angles $\theta$ and $\phi$ with $H_x=H\sin\theta\cos\phi$,
$H_y=H\sin\theta\sin\phi$, and $H_z=H\cos\theta$. In Fig.\,S5 we
present the  phase diagrams  for two field directions:  (a)  $\theta=\phi=0$,
{\it i.e.}, the field along the [0,0,1] axis; (b) $\theta=\phi=\pi/4$,  which
corresponds to a generic direction (as opposed to  high symmetry  directions).
At zero field, there is a continuous phase transition from the paramagnetic to
an AFQ ordered phase at $T_{\mathrm{Q}}\approx 0.28J_{\tau}$, but no magnetic dipolar order
is found down to zero temperature. When applying a magnetic field, along both
directions $T_{\mathrm{Q}}$ first rises up with increasing magnetic field, and the AFQ
phase is stable in a broad field range. At elevated fields, the phase diagram
shows strong anisotropy. The AFQ phase is more stable for the field direction
$\theta=\phi=\pi/4$ where it survives up to a higher temperature
($T\approx0.47J_\tau$), and to a  higher  magnetic field ($H\approx5.2J_\tau$).
The dominant AFQ moment is also different along the two directions. For
$\theta=\phi=0$, a $\Gamma_3$-type ($O_2^0$) AFQ order is stabilized at
intermediate fields, while the $\Gamma_5$-type order is stabilized at either low
or high fields. The regimes with different quadrupolar orders are separated by
first-order transitions. Along the $\theta=\phi=\pi/4$ direction, the
$\Gamma_3$- and $\Gamma_5$-type orders can coexist, but the dominant order
parameter shows a crossover from $O_{xz}+O_{yz}$ in low fields to $O_2^0$ in
high fields.

At zero magnetic field, the quadrupolar moments and the dipolar moments
transform according to different irreducible representations of the cubic group.
Therefore, no magnetic dipolar moment can be induced by the ordered quadrupolar
moment. A nonzero magnetic field reduces the cubic symmetry of the electronic
degrees of freedom, and the ordered AFQ moments may induce magnetic dipolar
moments with the same point group symmetry and same ordering wavevector. For
instance, the $O_2^0$-type AFQ order induces AFM moment $J^z$ under a magnetic
field along the [0,0,1] direction. Though there is no spontaneous magnetic
order, the induced AFM order will make part of or even the entire AFQ phase
magnetically active depending on the symmetry of the AFQ moments. These parts
are shown as the shaded areas in Fig.\,S5.

For a single crystalline sample in a magnetic field along a crystallographic
high symmetry direction, the AFQ ordered phase may not be entirely magnetically
ordered, as shown in the left panel of Fig.\,S5. However, the
situation is different for a polycrystalline sample. There, one has to average
over all the possible angles between the crystalline axes and the field. For any
generic field orientation, the AFQ ordering always induce magnetic order, as
exemplified by the right panel of Fig.\,S5. Therefore, for a
polycrystalline sample, the entire AFQ ordered phase will be AFM ordered. By
extension, such a material will not display any features associated with
first order transitions between AFQ orders.

{\noindent\it Ginzburg-Landau analysis}

Our microscopic studies show that the AFQ order can influence the magnetic
correlations by inducing AFM order with the same symmetry along certain magnetic
field directions. Here we consider the interplay of the AFQ order and the induced
symmetry-compatible AFM order via a Ginzburg-Landau theory. We show that if the
induced AFM moments are allowed by symmetry, they are immediately ordered at
$T_{\mathrm{Q}}$, and the induced AFM order enhances the AFQ order by increasing $T_{\mathrm{Q}}$ with
increasing magnetic field.

For simplicity, we assume that only one component of the quadrupolar moments is
ordered. The ordered quadrupole and its symmetry-compatible dipolar moments are
denoted as $\tau_{a(f)}$, $\sigma_{a(f)}$, and $\eta_{a(f)}$. For instance,
along the magnetic field direction $\theta=\phi=0$, the ordered moments are,
respectively, $O_2^0$ and $J^z$, which correspond to $\tau^z$, $\sigma^z$, and
$\eta^z$. (In the following discussion we will only focus on the interplay of
quadrupolar and dipolar moments though $\sigma^z$ and $\eta^z$ also include an
octupolar component with the same symmetry as the dipolar component.) We then
expand the Ginzburg-Landau free energy up to the fourth order of the pseudospin
operators
\begin{eqnarray}\label{Eq:Landau}
 \mathcal{F} = \mathcal{F}_0 &+& 
 \frac{1}{2} \sum_{\alpha=\tau,\sigma,\eta} \left[(T-T_0^\alpha) ( \phi^{\alpha}_a)^2 + (T+T_0^\alpha) (\phi^{\alpha}_f)^2 \right]
+r (\tau_f\sigma_f\eta_f + \tau_f\sigma_a\eta_a + \tau_a\sigma_f\eta_a + \tau_a\sigma_a\eta_f) \nonumber \\
&-& H_z\sum_{c=f,a} \tau_c(\frac{4}{7}\sigma_c+\eta_c) 
- H_z (\sigma_f  + \frac{4}{7} \eta_f) 
- (\delta H_\tau \tau_a +\delta H_\sigma \sigma_a +\delta H_\eta \eta_a) \nonumber \\
&+& \sum_{\phi=\tau,\sigma,\eta} \frac{u_\alpha}{4}(\phi^{\alpha}_a)^4 
+ \ldots
\end{eqnarray}
Here we have introduced staggered fields $\delta H$'s coupled to $\tau_a$,
$\sigma_a$, and $\eta_a$. For simplicity, in the free energy we neglect some
quartic terms which may be allowed by symmetry, such as $\sigma_a^2\tau_a^2$.
Including these terms will not change the results qualitatively. $T_0^\tau$ is
the ordering temperature for $\tau_a$ at zero magnetic field. $T_0^\sigma$ and
$T_0^\eta$ are ordering temperatures for magnetic dipolar moments in the absence
of AFQ ordering. Here we consider the situation $T_0^{\sigma(\eta)}\leqslant
T_0^\tau$, which is the case for Ce$_3$Pd$_{20}$Si$_6$. $u_\tau$, $u_\sigma$, and $u_\eta$ are
positive coupling strengths. Note that the first term in the second line in
equation~(\ref{Eq:Landau}) couples the quadrupolar moments not only to dipolar
moments, but also to the magnetic field.

We obtain a set of coupled non-linear equations for $\tau_{a(f)}$,
$\sigma_{a(f)}$, and $\eta_{a(f)}$ by minimizing $\mathcal{F}$ with respect to
these parameters. Solving these equations in the paramagnetic phase we have, to
the leading order in $H_z$
\begin{eqnarray}
 \sigma_f \approx \frac{H_z}{T+T_0^\sigma},\quad 
 \eta_f \approx \frac{4H_z}{7(T+T_0^\eta)},\quad \mathrm{and}\quad
 \tau_f \approx \frac{4H_z^2(2T+T_0^\sigma+T_0^\eta-r)}{7(T+T_0^\tau)(T+T_0^\sigma)(T+T_0^\eta)}.\label{Eq:ST15}
\end{eqnarray}
Now we investigate the paramagnetic to AFQ transition at a fixed $H_z$.
Linearizing the equations for $\tau_{a}$, $\sigma_{a}$, and $\eta_{a}$ (by
dropping the terms including $u$), we find that $\tau_a$, $\sigma_a$, and
$\eta_a$ become nonzero only when
\begin{eqnarray}
 \det\left|
 \begin{tabular}{ccc}
  $T-T_0^\tau$ & $r\eta_f-\frac{4}{7}H_z$ & $r\sigma_f-H_z$ \\
  $r\eta_f-\frac{4}{7}H_z$ & $T-T_0^\sigma$ & $r\tau_f$ \\
  $r\sigma_f-H_z$ & $r\tau_f$ & $T-T_0^\eta$
 \end{tabular}
\right|=0.\label{Eq:ST16}
\end{eqnarray}
This determines a single temperature $T_{\mathrm{Q}}$, at which the paramagnetic to AFQ
transition takes place in the presence of magnetic field. Solving
equation~(\ref{Eq:ST16}) after inserting in it equation~(\ref{Eq:ST15}), we
obtain $T_{\mathrm{Q}}\approx T_0^\tau+\frac{\sqrt{65}}{14}H_z$ for
$T_0^\tau=T_0^\sigma=T_0^\eta$, and $T_{\mathrm{Q}}\approx
T_0^\tau+\frac{65}{49}H_z^2(T_0^\tau+T_0-r)^2/[(T_0^\tau-T_0)(T_0^\tau+T_0)^2]$ 
for $T_0^\tau>T_0$, where $T_0=T_0^\sigma=T_0^\eta$. Hence we see that $T_{\mathrm{Q}}$
always increases with increasing  $H_z$ as long as $T_0^\tau\geqslant
T_0^{\sigma(\eta)}$.

This unconventional field dependence of $T_{\mathrm{Q}}$ is due to the coupling between the
quadrupolar and dipolar moments, which appears as the off-diagonal terms in
equation~(\ref{Eq:ST16}). The increase of $T_{\mathrm{Q}}$ with magnetic field is
consistent with the experimental phase diagram of Ce$_3$Pd$_{20}$Si$_6$. It is also compatible
with the microscopically-calculated phase diagram presented in Fig.\,S5.  It is
especially instructive to note that, in Fig.\,S5a, $T_{\mathrm{Q}}$ increases significantly
only in the $O_2^0$ phase, in which the AFM moment $J^z$ is induced. By
contrast, $T_{\mathrm{Q}}$ is almost unchanged in the $O_{xy}$ phase where no magnetic
moment can be induced. Note that though there is no induced dipolar moment in
the $O_{xy}$ phase along the $[0,0,1]$ direction, the $T_{xyz}$-type
antiferro-octupolar moment can be induced, which gives rise to the weak field
dependence of $T_{\mathrm{Q}}$ in this phase.

Equation~(\ref{Eq:ST16}) also implies that the magnetic dipolar moments
immediately become ordered at $T_{\mathrm{Q}}$, at which they are induced by the AFQ
order.  This conclusion survives fluctuation effects induced by the quartic
couplings. The latter will renormalize the diagonal and off-diagonal terms in
the  matrix appearing in equation~(\ref{Eq:ST16}), which will give rise to a
renormalized $T_{\mathrm{Q}}$.

{\noindent\it Effect of couplings to conduction electrons}

The consistency between our symmetry-based analysis and microscopic calculations
implies that the conclusions from the latter will survive any generalization of
the model that does not change its symmetry. One important generalization of
equation~(\ref{Eq:ST03}) is that the pseudospins will be Kondo-coupled to the
conduction electrons. Separately, the $\Gamma_7$ doublets of the Ce ions at the
$4a$ site will also be Kondo-coupled to the conduction electrons. Through the
conduction electrons, the two sets of the Ce 4$f$ electrons will be coupled to
each other in the dipolar channel.  These interactions will generate a Kondo
screened ground state in the absence of a competition by RKKY interactions. The
latter  comprise interactions within the $4a$ and $8c$ sites, respectively,
which are responsible for the  AFQ and AFM orders, as well as those  between
$4a$ and $8c$ sites, which couple the AFM orders of the two components. The
competition between the RKKY and Kondo couplings gives rise to the Kondo
destruction.

%FIGURE LEGENDS AND TABLE SUPPLEMENTARY INFORMATION !!!!!!!!!!!!!!!!!!!!!!!!!

\newpage

%\begin{figure}[ht!]
%\caption{%\label{figS1}
\noindent FIG.\,S1: Non-Fermi liquid behaviour of Ce$_3$Pd$_{20}$Si$_6$.
{\bf{a}} Difference of electrical resistivity $\rho(T)$ and its residual
resistivity $\rho_0$ vs temperature, at different magnetic fields. {\bf{b}}
Electronic specific heat coefficient $\Delta C/T$ vs temperature on a
semi-logarithmic scale for a field of 1~T (main panel) and for several fields
(inset). Both the linear $T$ dependencies in $\rho(T)$ and the
$-\ln{T}$ dependencies in $\Delta C/T$ are indicated by dashed lines.\\
%}
%\end{figure}

%\begin{figure}[ht!]
%\caption{%\label{figS2}
\noindent FIG.\,S2: Fitting procedure for magnetoresistivity isotherms. Field dependence
at 300~mK of the longitudinal resistivity $\rho_{\mathrm{l}}$, corrected for the
background contribution $\rho_{\mathrm{l,back}}$, with the sum of the two
crossover functions given in the figure. The crossover positions
$B_{\mathrm{N}}$ and $B^{\ast}$ are indicated.\\
%}
%\end{figure}

%\begin{figure}[ht!]
%\caption{%\label{figS3}
\noindent FIG.\,S3: Characteristics of the crossover at the N\'eel transition. Temperature
dependence of the step height $\Delta A_{\mathrm{N}}$ in {\bf a} and of the
crossover width $w_{\mathrm{N}}$ in {\bf b}, determined from fits to the
longitudinal and transverse resistivity $\rho_{\mathrm{l}}$ and
$\rho_{\mathrm{t}}$, both corrected for the background contribution, and to the
Hall resistivity $\rho_{\mathrm{H}}$. The star refers to Hall resistivity data
corrected for the anomalous Hall effect.\\
%}
%\end{figure}

%\begin{figure}[ht!]
%\caption{%\label{figS4}
\noindent FIG.\ S4: Analysis of the anomalous contribution to the Hall
resistivity. Measured Hall resistivity $\rho_{\mathrm{H}}$ (symbols) and fits to
the data (lines through the data points), as well as the normal contribution
$\rho_{\mathrm{H}}^0$ estimated from magnetization and resistivity data as
discussed in the text, vs ratio of magnetic field and crossover field
$B/B^{\ast}$.\\
%}
%\end{figure}

%\begin{figure}[ht!]
%\caption{%\label{Fig:S5}
\noindent FIG.\ S5: Finite-temperature mean-field phase diagram. The parameter
set is given in the text. The external magnetic field is along the directions
$\theta=\phi=0$ in {\bf a} and $\theta=\phi=\pi/4$ in {\bf b}. The black solid
curve ($T_{\mathrm{Q}}$) shows a continuous phase transition between the paramagnetic and
the AFQ ordered phase. In {\bf a} the blue dash dotted lines represent
first-order transitions inside the AFQ phase, while in {\bf b} the blue
dash-dotted line indicates a crossover. The shaded areas refer to the
magnetically ordered regimes.\\
%}
%\end{figure}

\begin{table}
\caption{Pseudospin representation of multipolar moments, adapted from
ref.\,\onlinecite{Shiina97}. $\mathbf{J}$, $O$, and $\mathbf{T}$ correspond to
dipolar, quadrupolar, and octupolar operators, respectively. Also listed are the
corresponding irreducible representations of the $O_h$ point
group.}\label{TableSTI}
\vspace{0.5cm}

 \begin{tabular}{cccc}
  \hline\hline
	pseudospin operator & $(\boldsymbol\tau,\boldsymbol\sigma)$ representation &
multipolar moment & symmetry\\
  \hline
  $\boldsymbol\tau^\prime$ & $(\tau^z,\tau^x)$ & $\frac{1}{8}(O_2^0,O_2^2)$ & $\Gamma_3^+$ \\
  $\boldsymbol\mu$ & $2(\tau^y\sigma^x,\tau^y\sigma^y,\tau^y\sigma^z)$ & $\frac{1}{2}(O_{yz},O_{xz},O_{xy})$ & $\Gamma_5^+$ \\
  $\boldsymbol\sigma$ & $2(\sigma^x,\sigma^y,\sigma^z)$ & $\frac{7}{15}\mathbf{J}-\frac{2}{45}\mathbf{T}^\alpha$ & $\Gamma_4^-$ \\
  $\boldsymbol\eta$ & $(-\tau^z\sigma^x+\sqrt{3}\tau^x\sigma^x,-\tau^z\sigma^y-\sqrt{3}\tau^x\sigma^y,2\tau^z\sigma^z)$ & $-\frac{1}{15}\mathbf{J}+\frac{7}{90}\mathbf{T}^\alpha$ & $\Gamma_4^-$ \\
  $\boldsymbol\zeta$ & $(-\sqrt{3}\tau^z\sigma^x-\tau^x\sigma^x,\sqrt{3}\tau^z\sigma^y-\tau^x\sigma^y,2\tau^x\sigma^z)$ & $\frac{\sqrt{5}}{30}\mathbf{T}^\beta$ & $\Gamma_5^-$ \\
  $\xi$ & $\tau^y$ & $\frac{\sqrt{5}}{45}T_{xyz}$ & $\Gamma_2^-$ \\
  \hline
 \end{tabular}
\vspace{10cm}
\end{table}

%FIGURES SUPPLEMENTARY INFORMATION !!!!!!!!!!!!!!!!!!!!!!!!!!!!!!!!!

\newpage

\begin{figure}[t!]
\centerline{\includegraphics[width=110mm]{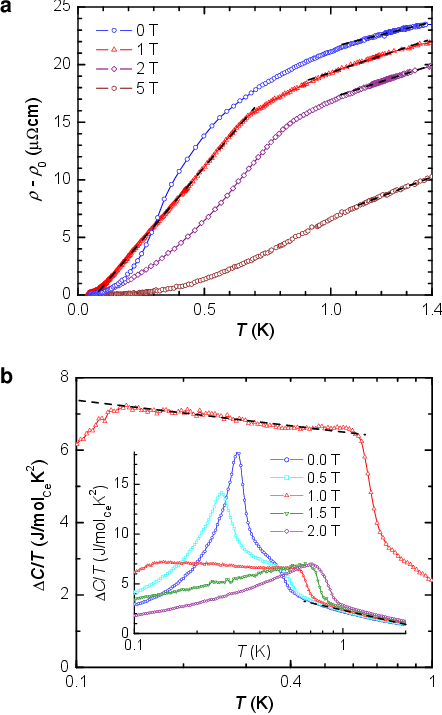}}
{\Large Figure S1, Supplementary Information}
\end{figure}

\vspace{2cm}

\newpage

\begin{figure}[t!]
\centerline{\includegraphics[width=110mm]{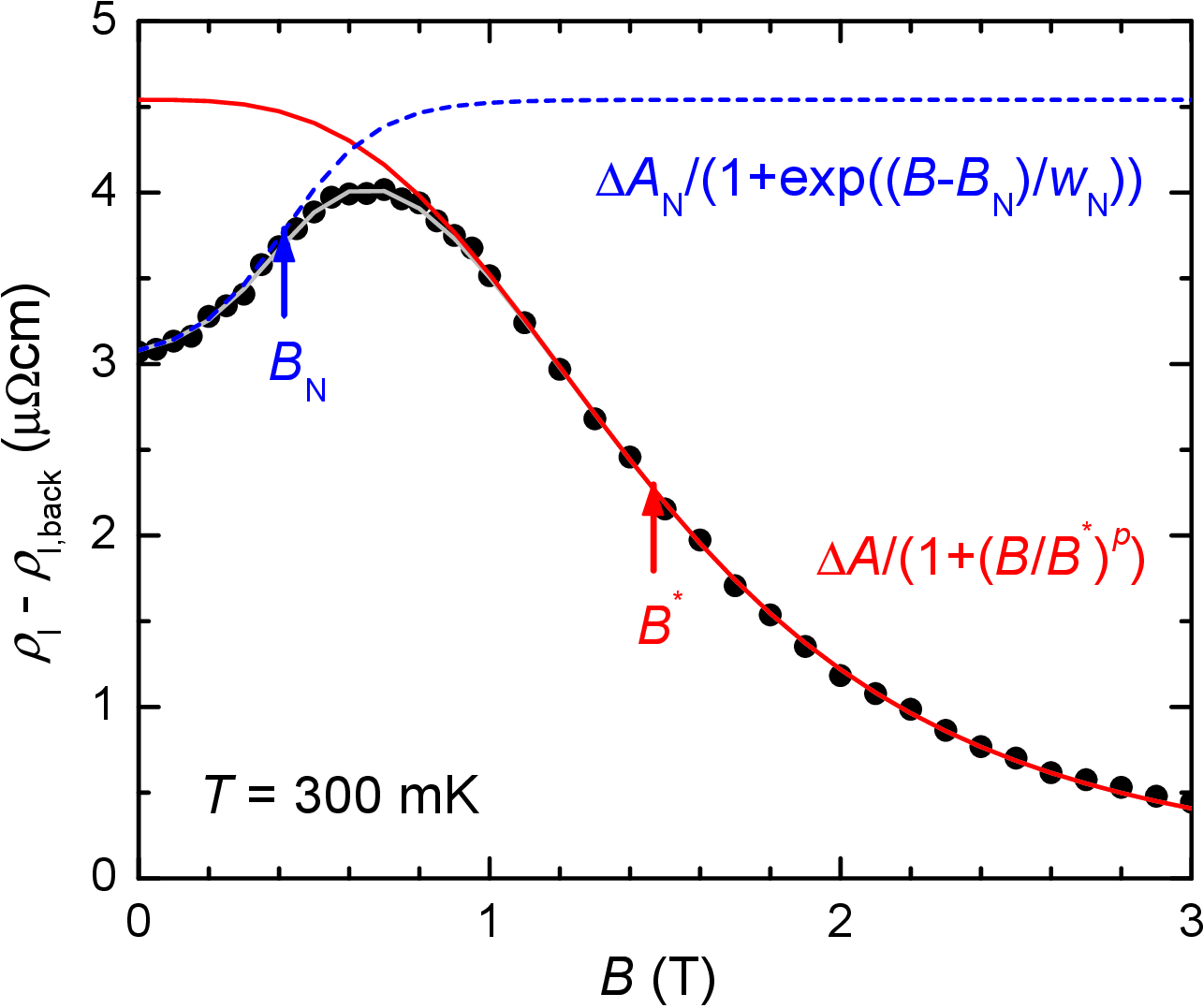}}
{\Large Figure S2, Supplementary Information}
\end{figure}

\vspace{2cm}

\begin{figure}[t!]
\centerline{\includegraphics[width=110mm]{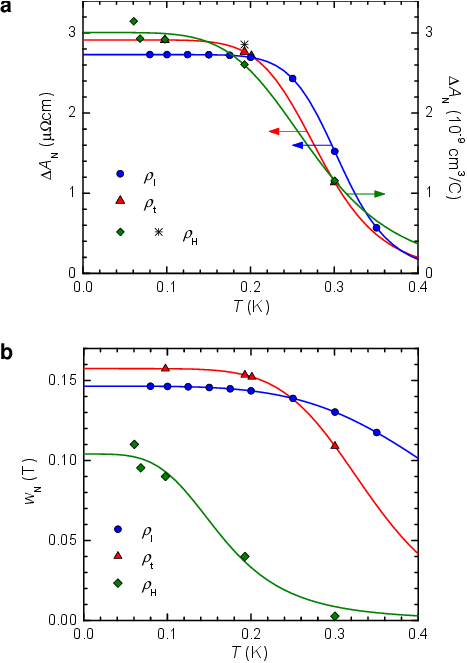}}
{\Large Figure S3, Supplementary Information}
\end{figure}

\begin{figure}[b!]
\centerline{\includegraphics[width=110mm]{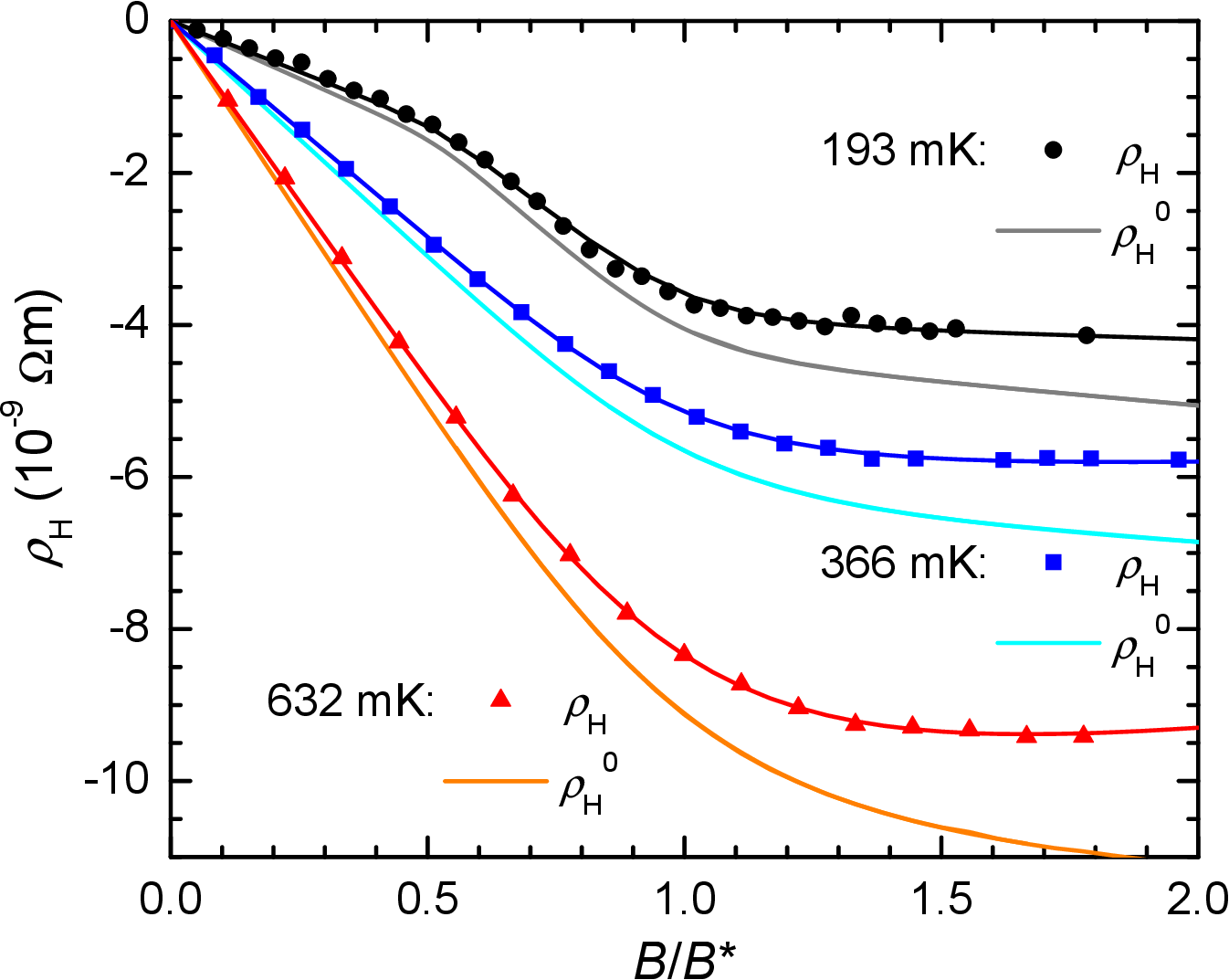}}
{\Large Figure S4, Supplementary Information}
\end{figure}

\newpage

\begin{figure}[t!]
\centerline{\includegraphics[width=110mm]{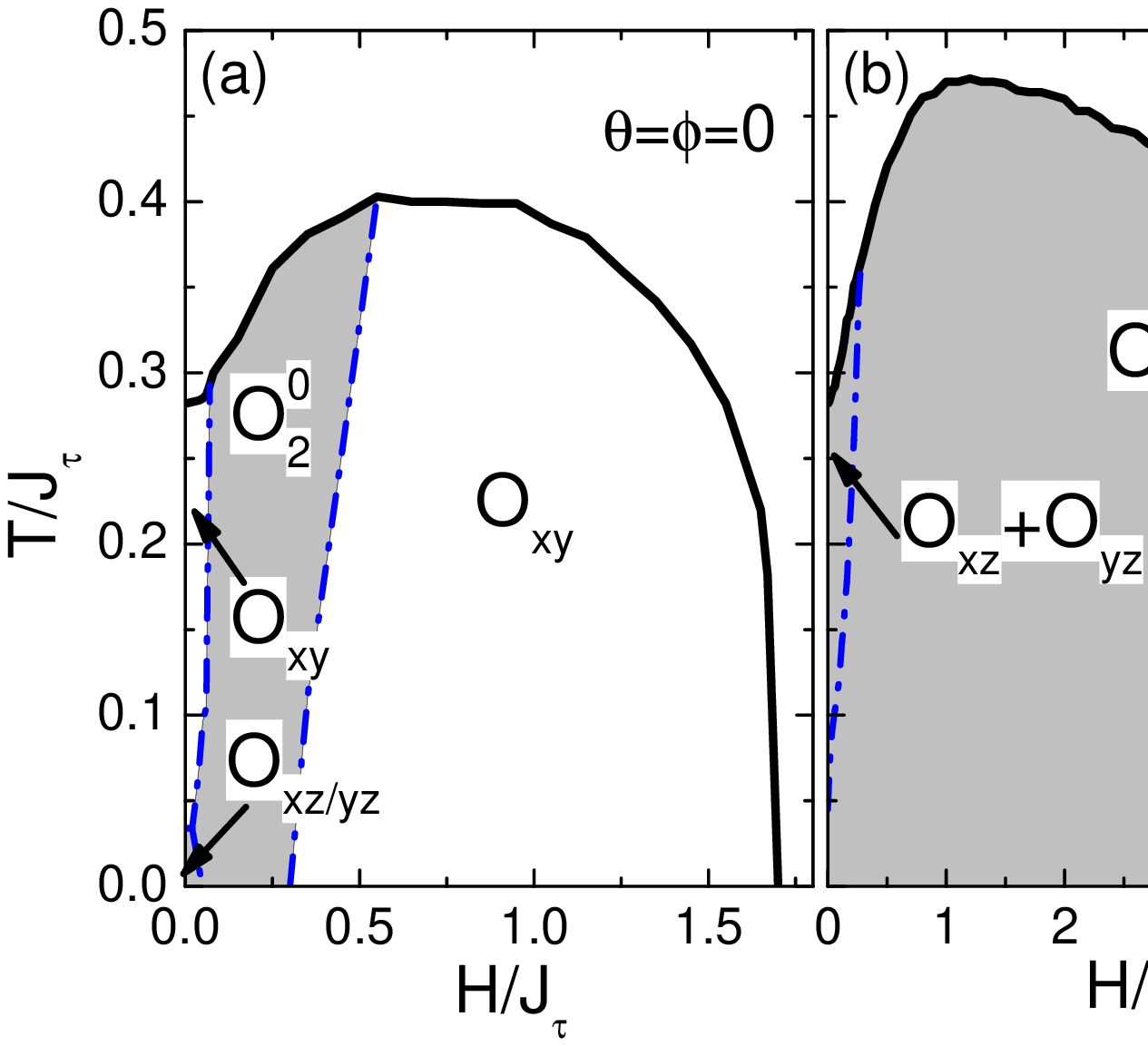}}
{\Large Figure S5, Supplementary Information}
\end{figure}